\documentclass[conference]{IEEEtran}
\IEEEoverridecommandlockouts
\usepackage{cite}
\usepackage{amsmath,amssymb,amsfonts}
\usepackage{algorithmic}
\usepackage{graphicx}
\usepackage{textcomp}
\usepackage{xcolor}
\usepackage{ucs}
\usepackage[utf8x]{inputenc}
\usepackage{cite, amsfonts, amssymb, amsthm, bm, bbm, graphicx, relsize, multirow, booktabs, tikz,subfigure,soul}
\usepackage{stackrel}
\usepackage{tabulary}
\usepackage[american]{babel}
\usepackage[T1]{fontenc}
\usepackage{algorithmic, algorithm}
\usepackage[multiple]{footmisc}

\newcommand{\ds}{\displaystyle}

\def\BibTeX{{\rm B\kern-.05em{\sc i\kern-.025em b}\kern-.08em
    T\kern-.1667em\lower.7ex\hbox{E}\kern-.125emX}}
\begin{document}
% comando per disabilitare i dashes quando i nomi degli autori si ripetono (IEEEexample:BSTcontrol definito in finalRefs.bib)
\bstctlcite{IEEEexample:BSTcontrol}
\title{User Association in Scalable Cell-Free Massive MIMO Systems}

\author{\IEEEauthorblockN{Carmen D'Andrea\IEEEauthorrefmark{1} and Erik G. Larsson\IEEEauthorrefmark{2}}
\IEEEauthorblockA{\IEEEauthorrefmark{1}\textit{DIEI - University of Cassino and Southern Latium},
Cassino, Italy.}
\IEEEauthorblockA{\IEEEauthorrefmark{2} \textit{ISY - Link\"{o}ping University}, 
Link\"{o}ping, Sweden.}
\thanks{This work was partly developed during the visiting period of the first author to Link\"{o}ping University, Link\"{o}ping, Sweden. }}

\maketitle

\begin{abstract}
In this work, we consider the uplink of a scalable cell-free massive MIMO system where the users are served only by a subset of access points (APs) in the network. The APs are physically grouped into predetermined ``cell-centric clusters'', which are connected to different cooperative central processing units (CPUs). Given the cooperative nature of the considered communications network, we assume that each user is associated with a ``virtual cluster'', that, in general, involves some APs belonging to different cell-centric clusters. Assuming the maximum-ratio-combining at the APs, we propose a user-association procedure aimed at the maximization of the sum-rate of the users in the system. The proposed procedure is based on the Hungarian Algorithm and exploits only the knowledge of the position of the APs in the network. Numerical results reveal that the performance of the proposed approach is not always better than the alternatives but it offers a considerably lower backhaul load with a negligible performance loss compared to full-cell free approaches.

\end{abstract}

\begin{IEEEkeywords}
cell-free massive MIMO, user association, scalability, Hungarian algorithm.
\end{IEEEkeywords}

\section{Introduction}
The cell-free massive MIMO technology consists of a very large number of distributed APs serving many users in the same time-frequency resource \cite{ngo2015cell,Ngo_CellFree2017}. In the originally formulated version of these systems, all the APs serve all the users in the coverage area, i.e., there are no cells or cell boundaries. All the APs are connected to a common CPU, where the data encoding/decoding is performed. Going on with the research on this topic, the hypothesis in which the users are served by all the APs in the network has been relaxed. One kind of approach is the user-centric (UC) assignment in which each user is served only by a subset of APs in the system\cite{buzzi_CFUC2017,Buzzi_DAndrea_Zappone_TWC2019}. In this approach, the concept of traditional \textit{cell} has been overcome, in the sense that there is no longer the cell defined starting from the network topology, but it is the network that is getting closer to users. These UC assignment, particularly when the quality of the channel estimation is low and the simple maximum-ratio processing is implemented at the APs, offers better performance with respect to the case in which all the APs serve all the users in the system and gives at least two benefits. On one hand, serving the each user by a subset of APs allows us to better focus the available power at each AP on the ``useful'' transmissions without wasting power for the one that are in very bad channel conditions. On the other hand, the UC assignment reduces the load on the backhaul link, because, for each user, only a subset of APs are involved in the data encoding/decoding and shares the information of that user on the backhaul link. Another kind of approach that improves the performance of the cell-free massive MIMO systems over the baseline maximum-ratio processing is the large-scale fading decoding (LSFD) \cite{nayebi2016performance}. In the LSFD, the users are served by all the APs in the network but data are weighted at the CPU based on the knowledge of the large-scale fading (LSF) coefficients between all the APs and the users. The main problem in the cell-free massive MIMO systems is the scalability, i.e., when the number of users increases the complexity of the system becomes higher and higher because the main processing capabilities are entrusted to the common CPU. A scalable version of the cell-free massive MIMO is proposed and discussed on the downlink in \cite{Interdonato_Scalability2019} considering data transmission and power control strategies where multiple cooperative CPUs serve disjoint clusters of APs.

In this work, we focus on the uplink of a scalable cell-free massive MIMO system considering a UC approach, in which the association between users and APs is performed maximizing the uplink sum-rate of the whole system. In particular, exploiting the uplink spectral efficiency lower bound that depends only on the subset of the APs decoding each user, we formulate the association problem as a matching problem that can be solved using the Hungarian algorithm \cite{kuhn1955hungarian}. Numerical results reveal the superiority of the proposed approach, both in terms of sum-rate and in terms of rate per user with respect to the full cell-free approach in \cite{ngo2015cell,Ngo_CellFree2017} and to the UC approach where each user is served by the APs with the best channels in \cite{buzzi_CFUC2017,Buzzi_DAndrea_Zappone_TWC2019}. Additionally, numerical results reveal that the performance of the proposed approach is not always superior in terms of performance than the LSFD, but it offers a considerably lower backhaul load with a negligible performance loss with respect to the full cell-free system with an additional processing at the CPU exploiting the LSF coefficients.

%\subsection{Notation}
%We use non-bold letters for scalars, $a$ and $A$, lowercase boldface letters, $\mathbf{a}$, for vectors and uppercase lowercase letters, $\mathbf{A}$, for matrices. The transpose, the inverse and the conjugate transpose of a matrix $\mathbf{A}$ are denoted by $\mathbf{A}^T$, $\mathbf{A}^{-1}$ and $\mathbf{A}^H$, respectively. The $N$-dimensional identity matrix is denoted as $\mathbf{I}_N$, the $(N \times M)$-dimensional matrix with all zero entries is denoted as $\mathbf{0}_{N \times M}$ and $\mathbf{1}_{N \times M} $ denotes a $(N \times M)$-dimensional matrix with unit entries. The statistical expectation operator is denoted as $\mathbb{E}[\cdot]$; $\mathcal{CN}\left(\mu,\sigma^2\right)$ denotes a complex circularly symmetric Gaussian random variable with mean $\mu$ and variance $\sigma^2$.

\section{System model}

\begin{figure}[!t]
\centering
\includegraphics[scale=0.28]{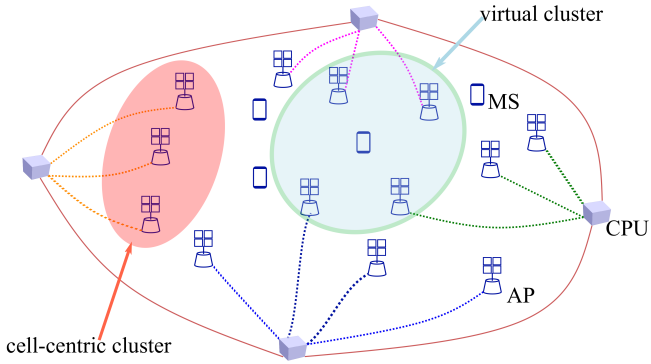}
\caption{Cell-free massive MIMO system with virtual clustering and multiple interconnected CPUs.}
\label{Fig:CF_MIMO}
\end{figure}

We consider a scalable implementation of cell-free massive MIMO system \cite{Interdonato_Scalability2019} reported in Fig. \ref{Fig:CF_MIMO} with $M$ APs with $N_{\rm AP}$ antennas and $K$ single-antenna users. In particular, the network consists of outdoor APs and users and the APs are grouped into $N$ predetermined \textit{cell-centric clusters}. The indexes of the APs belonging to the same cell-centric cluster are contained in the sets $\mathcal{C}_1,\ldots,\mathcal{C}_N$. Each cell-centric cluster in turn is connected to one \textit{primary CPU}, the primary CPUs are interconnected but operate autonomously. It is assumed that a global phase reference is shared to allow synchronization of all the actors in the communications system. In keeping with the approach in \cite{ngo2015cell,Ngo_CellFree2017}, all communications take place on the same frequency band, i.e. uplink and downlink are separated through TDD.
The generic $k$-th MS is served by a subset of the APs on a given physical resource block (PRB), we call this subset \textit{virtual cluster} (VC) and it is determined based on a ``neighbourhood criterion'' between the APs. We denote by $\mathcal{M}_1, \ldots, \mathcal{M}_J$ the sets containing the indexes of the APs in each VC. Note that, while the cell-centric clusters are disjoint, because connected to different CPUs, the VCs can be overlapped in order to allow the cooperation between the primary CPUs to serve the users in the system. 

We denote by the $N_{\rm AP}$-dimensional vector $\mathbf{g}_{k,m}$ the channel between the $k$-th MS and the $m$-th AP written as

\begin{equation}
\mathbf{g}_{k,m}=\sqrt{\beta_{k,m}}\mathbf{h}_{k,m} ,
\label{channel_model}
\end{equation}
with $\beta_{k,m}$ a scalar coefficient modeling the channel LSF effects and 
$\mathbf{h}_{k,m}$ an $N_{\rm AP}$-dimensional vector whose entries are i.i.d ${\cal CN}(0,1)$ random variables (RVs) modelling the fast fading.
\vspace{-0.5cm}
\subsection{Uplink Training}
\label{Ch_est}

The dimension in time/frequency samples of the channel coherence length is denoted by $\tau_c$, and the dimension of the uplink training phase by $\tau_p < \tau_c$. The pilot sequence transmitted by the generic $k$-th MS, say $\boldsymbol{\phi}_k$, is chosen from a set of $\tau_p$ orthogonal sequences $\mathcal{P}_{\tau_p}=\left\lbrace \boldsymbol{\psi}_1, \boldsymbol{\psi}_2, \ldots, \boldsymbol{\psi}_{\tau_p} \right\rbrace$, where $\boldsymbol{\psi}_i$ is the $i$-th $\tau_p$-dimensional column sequence and $\|\boldsymbol{\psi}_i\|^2=1$, $ \forall \, i=1,\ldots, \tau_p$.

Exploiting the knowledge of the users' pilot sequences, the $m$-th AP can estimate the channel vectors $\mathbf{g}_{k,m}$ by projecting the received signal on the pilot sequence of the $k$-th user, i.e., it forms the statistics
\begin{equation}
\begin{array}{llll}
\widehat{\mathbf{y}}_{k,m}=\mathbf{Y}_m \boldsymbol{\phi}_k=& \sqrt{p_k}\mathbf{g}_{k,m} \\ & + \ds \sum_{\substack{i=1 \\ i\neq k}}^K {\sqrt{p_i}\mathbf{g}_{i,m}\boldsymbol{\phi}_i^H \boldsymbol{\phi}_k} + \mathbf{W}_m \boldsymbol{\phi}_k \; ,
\label{y_hat_ka}
\end{array}
\end{equation}
where ${p}_k=\tau_p \widetilde{p}_k$ denotes the power employed by the $k$-th user during the training phase, and $\mathbf{W}_m$ contains the thermal noise contribution at the $m$-th AP, with i.i.d. ${\cal CN}(0, \sigma^2_w)$ entries.

Assuming knowledge of the LSF coefficients $\beta_{k,m}, \, \forall \; k,m$, the minimum-mean-square-error (MMSE) channel estimate of the channel $\mathbf{g}_{k,m}$ can be written as
\begin{equation}
\widehat{\mathbf{g}}_{k,m}= \frac{\sqrt{p_k}\beta_{k,m}}{\ds \sum_{i=1}^K p_i \beta_{i,m} \left|\boldsymbol{\phi}_i^H \boldsymbol{\phi}_k\right|^2 +\sigma^2_w} \widehat{\mathbf{y}}_{k,m} = \alpha_{k,m} \widehat{\mathbf{y}}_{k,m} \; .
\label{LMMSE_est2}
\end{equation}
The channel estimation error is given by $\widetilde{\mathbf{g}}_{k,m}=\mathbf{g}_{k,m} - \widehat{\mathbf{g}}_{k,m}$, and the estimate and the estimation error are independent \cite{kay1993fundamentals}. They are distributed as $\widehat{\mathbf{g}}_{k,m} \sim \mathcal{CN}\left(\mathbf{0}_{N_{\rm AP}}, \gamma_{k,m} \mathbf{I}_{N_{\rm AP}}\right)$, and $\widetilde{\mathbf{g}}_{k,m} \sim \mathcal{CN}\left(\mathbf{0}_{N_{\rm AP}}, (\beta_{k,m}-\gamma_{k,m}) \mathbf{I}_{N_{\rm AP}}\right)$, respectively, where $\gamma_{k,m}$ is the mean-square of the estimate, i.e., $ \gamma_{k,m}=\mathbb{E} \left[ \left|\left(\widehat{\mathbf{g}}_{k,m}\right)_{\ell}\right|^2 \right]= \sqrt{p_k}\beta_{k,m}\alpha_{k,m} \, ,\ell=1,\ldots, N_{\rm AP}$.
\subsection{Uplink Data Transmission} 
\label{UL_Section}
In uplink, users send their data symbols without any channel-dependent phase offset. As a result, the signal $\bar{\mathbf{y}}_m$ received at the $m$-th AP in the generic symbol interval is 
\begin{equation}
{\bar{\mathbf{y}}}_m=\ds \sum_{k=1}^K \ds \sqrt{\eta_{k}} \mathbf{g}_{k,m} {x}_k + \mathbf{w}_m \; ,
\end{equation}
with ${\eta_{k}}$ and ${x}_k$ representing the uplink transmit power and the data of the $k$-th user in the generic symbol interval, respectively, and $\mathbf{w}_m \sim {\cal CN}(\mathbf{0}, \sigma^2_w \mathbf{I} )$ the noise vector.

Subsequently, the $m$-th AP decodes the data transmitted by a subset of users in the system, say ${\cal K}_m$, and forms, for each $k \in {\cal K}_m$,  the  statistics
${{t}}_{m,k}= \mathbf{v}_{k,m}^H {\bar{\mathbf{y}}}_m(n)$, with $\mathbf{v}_{k,m}$ the combining vector for the $k$-th user, and sends them to its primary CPU. We assume that the $k$-th user is assigned to the VC whose AP indexes are in the set $\mathcal{M}_j$, determined according to such assignment criterion that will be specified later. We denote by $\mathcal{D}_{k,n}$ the set containing the APs belonging to the $n$-th cell-centric cluster that decode the $k$-th MS, i.e., 
\begin{equation}
\mathcal{D}_{k,n}= \{ m \in \mathcal{C}_n \; : \; k \in \mathcal{K}_m  \} \, .
\label{Def_D_kn}
\end{equation}
If the set $\mathcal{D}_{k,n}$ is not empty, i.e., at least one AP in $\mathcal{C}_n$ decodes the $k$-th user, the $n$-th primary CPU shares the following statistic with the other CPUs:
\begin{equation}
\widetilde{{x}}_{k,n} = \ds \sum_{m \in \mathcal{D}_{k,n}} {{t}}_{m,k} \; , \quad k  \; : \; \mathcal{D}_{k,n} \neq \emptyset \, .
\label{Est_UL_uc1_nCPU}
\end{equation}
We define $\mathcal{B}_k$ as the set containing the indexes of the primary CPUs that cooperate to decode the $k$-th MS, i.e.,
\begin{equation}
\mathcal{B}_{k}= \{ n  : \; \mathcal{D}_{k,n} \neq \emptyset  \} \, .
\label{Def_B_k}
\end{equation}
Finally, the soft estimation of the signal transmitted by the $k$-th MS can be written as
\begin{equation}
\begin{array}{llll}
\widehat{{x}}_{k} &= \ds \sum_{n \in \mathcal{B}_{k}} \widetilde{{x}}_{k,n}, \; \forall \; k=1,\ldots, K.
\end{array}
\label{Est_UL_uc1_MS}
\end{equation}
Using the definitions in Eqs. \eqref{Def_D_kn}, \eqref{Def_B_k} and \eqref{Est_UL_uc1_MS}, the signal $\widehat{{x}}_{k}$ can be written as
\begin{equation}
\begin{array}{llll}
\widehat{x}_k = & \ds \sum_{m \in \mathcal{M}_j} {{t}}_{m,k}= \ds \sum_{m\in{\cal M}_j} \ds \sqrt{\eta_{k}} \mathbf{v}_{k,m}^H  \mathbf{g}_{k,m} {x}_k \\ & +
\ds \sum_{\substack{\ell=1 \\ \ell \neq k}}^K \ds \sum_{m \in \mathcal{M}_j} \sqrt{\eta_{\ell}}\mathbf{v}_{k,m}^H   \mathbf{g}_{\ell,m} {x}_{\ell}+  
 \ds \sum_{m\in{\cal M}_j} {\mathbf{v}_{k,m}^H \mathbf{w}_m } .
 \end{array}
\label{Est_UL_uc1_MS2}
\end{equation}
The meaning of the main mathematical symbols used in the system model can be found in Table \ref{Symbols_table}.

\begin{table}
\centering
\caption{Meaning of the main mathematical symbols}
\label{Symbols_table}
\def\arraystretch{1.2}
\begin{tabulary}{\columnwidth}{ |p{3cm}|p{4.5cm}| }
\hline
  \textbf{Symbols} 				& \textbf{Meanings} \\
\hline
  $M, K, N_{\rm AP}$ 				& Numbers of APs, users and antennas at each AP\\ \hline
  $\mathcal{C}_1,\ldots,\mathcal{C}_N$				& Sets of APs in the cell-centric clusters\\ \hline
  $\mathcal{M}_1,\ldots,\mathcal{M}_J$				& Sets of APs in the VCs\\ \hline
	$\mathcal{K}_m$		&  Set containing the users assigned to the $m$-th AP \\ \hline
	$\mathcal{D}_{k,n}$			& Set containing the APs in the $n$-th cell-centric cluster that decode the $k$-th user\\ \hline
	$\widetilde{x}_{k,n}$			& local estimate of the symbol transmitted by the $k$-th user computed at the $n$-th primary CPU\\ \hline
	$\mathcal{B}_k$ 		& Set containing the primary CPUs that cooperate to decode the $k$-th user\\ \hline
\end{tabulary}
\end{table}

\section{Performance measure and User Association rule}
We consider the uplink sum-rate of the users in the system as performance measure for the proposed association rule. Assuming that the CPUs rely only on statistical knowledge of the channel coefficients when performing the detection, according to the use-and-then-forget bounding technique, we model the sum of the beamforming uncertainty, interference and filtered thermal noise as ``effective noise'' and use the worst-case Gaussian assumption as in \cite{marzetta2016fundamentals,Ngo_CellFree2017}, which leads to the following lower-bound for the uplink achievable rate of the $k$-th when it is assigned to the APs in the set $\mathcal{M}_j$ as 
\begin{equation}
\mathcal{R}_{k}^{(j)}= \frac{\tau_{\rm u}}{\tau_c} W \log_2 \left( 1  +  \text{SINR}_k^{(j)} \right),
\label{eq:SE_general_UL}
\end{equation} 
where $\tau_{\rm u}$ is the length (in time-frequency samples) of the uplink data transmission phase in each coherence interval, $W$ is the system bandwidth. The signal-to-interference-noise ratio (SINR) $\text{SINR}_k^{(j)}$ is evaluated in the case of maximum ratio combining (MRC), i.e., $\mathbf{v}_{k,m}=\widehat{\mathbf{g}}_{k,m}$, and MMSE channel estimation and obtained in closed form, see Eq. \eqref{eq:SINR_bar_UL_LMMSE} at the top of next page, using similar derivations as in \cite{Ngo_CellFree2017}.

\begin{figure*}
\begin{equation}
\!\!\!\!\!\!\!\!\!\!\!\!\text{SINR}_{k}^{(j)} =\frac{ \eta_{k}  N_{\rm AP} \left( \ds \sum_{m\in{\cal M}_j} {\ds  \gamma_{k,m}} \right)^2}{
\ds \sum_{\ell=1}^K \eta_{\ell} \sum_{m\in{\cal M}_j} \beta_{\ell,m} \gamma_{k,m} + \ds \sum_{ \substack{\ell=1 \\ \ell\neq k}}^K \eta_{\ell} N_{\rm AP} \left( \ds \sum_{m \in {\cal M}_j}  \gamma_{k,m} \frac{\beta_{\ell,m}}{\beta_{k,m}}\right)^2  \left|\boldsymbol{\phi}_{\ell}^H \boldsymbol{\phi}_k \right|^2 + \sigma^2_w \!\!\!\!\sum_{m\in{\cal M}_j} {\!\! \gamma_{k,m}}
} 
\label{eq:SINR_bar_UL_LMMSE}
\end{equation}
\hrulefill
\end{figure*}
Note that in Eq. \eqref{eq:SINR_bar_UL_LMMSE} all the users contribute to interference in the sums over $\ell$ in the denominator, but the interference and the useful signal are only ``absorbed'' through the access points in $\mathcal{M}_j$ and not through all the APs in the network as in reference \cite{Ngo_CellFree2017}.
In the design of the proposed user association rule, we note that only the APs that decode the $k$-th user appear in the uplink SINR in Eq. \eqref{eq:SINR_bar_UL_LMMSE}.
Consequently, we formulate the following matching problem for the user association:
\begin{subequations}\label{Prob:Matching_problem1}
\begin{align}
\ds\max_{z_{k,j}}\;&\ds \sum_{k=1}^{K}\sum_{j=1}^{J} z_{k,j} \log_2 \left( 1 +  \text{SINR}_{k}^{(j)} \right) \label{Prob:aMatching_problem1}\\
\;\textrm{s.t.}\;& \sum_{k=1}^{K} z_{k,j} =1 \; \forall \, j=1, \ldots ,J\label{Prob:bMatching_problem1} \\
 \; \;& z_{k,j} \in \left\lbrace 0, 1\right\rbrace \; \forall \, k, j , \label{Prob:dMatching_problem1} 
\end{align}
\end{subequations} 
where we are neglecting the multiplicative constants in Eq. \eqref{eq:SE_general_UL} and we are looking for the optimal assignment aimed at the maximization of the uplink sum-rate, with the constraint according to each user is assigned to one of the $J$ predetermined VCs, $\mathcal{M}_1, \ldots, \mathcal{M}_J$.
This matching problem can be solved optimally in polynomial time by applying the Hungarian algorithm \cite[Algorithm 14.2.3]{jungnickel2007graphs}. This method is one of of the best known and most important combinatorial algorithms used to solve weighted matching problem in a bipartite graph. This algorithm is due to Harold Kuhn \cite{kuhn1955hungarian} and is based on ideas of two Hungarian mathematicians K\"onig and Egerv\`ary, so that Kuhn named it the \textit{Hungarian algorithm}. The application of the Hungarian algorithm for the solution of resource allocation problems in cell-free massive MIMO systems can be also found in reference \cite{BuzziDAndrea_Pilot_WCL2020}.

In Algorithm \ref{Hun_method}, we report a fast and efficient implementation of the Hungarian algorithm introduced by James Munkres in \cite{munkres1957algorithms} starting from a $(K \times J)$-dimensional weight matrix  $\mathbf{F}$ whose $(k,j)$-entries are $
f_{\left(k,j\right)}=\log_2 \left( 1 +  \text{SINR}_{k}^{(j)} \right) \, .
$

\begin{algorithm}[!t]

\caption{Hungarian algorithm \cite{kuhn1955hungarian,munkres1957algorithms,jungnickel2007graphs}}

\begin{algorithmic}[1]

\label{Hun_method}
\STATE  Find the maximum value in the matrix $\mathbf{F}$, say $f^+$.

\STATE Compute $\widetilde{\mathbf{F}}=f^+\mathbf{1}_{K \times J} - \mathbf{F}$. 

\STATE For each row of $\widetilde{\mathbf{F}}$, find the lowest element and subtract it from each element in that row.
 
\STATE For each column of the resulting matrix, find the lowest element and subtract it from each element in that column.

\STATE Cover all zeros in the resulting matrix using a minimum number of horizontal and vertical lines and denote as $\ell$ the number of total lines.

\IF{$\ell=K$}

\STATE An optimal assignment exists among the zeros. The algorithm stops.

\ELSE

\REPEAT

\STATE Find the smallest element, say $\widetilde{f}^*$, that is not covered by a line. 

\STATE Subtract $\widetilde{f}^*$ from all uncovered elements, and add $\widetilde{f}^*$ to all elements that are covered twice.

\STATE Cover all zeros in the resulting matrix using a minimum number of horizontal and vertical lines and denote as $\ell$ the number of total lines.

\UNTIL {$\ell=K$}

\STATE An optimal assignment exists among the zeros. The algorithm stops.

\ENDIF

\end{algorithmic}

\end{algorithm}
 
\subsection{Virtual Clustering}
In this paper, we propose a virtual clustering approach based on the positions of the APs in the network, that we call \textit{position-based virtual clustering} (PBVC). Note that, the information on the positions of the APs is reasonably available to the network operator that physically deploys the radiating resources to cover a given area. In this approach, each AP collects the indexes of the $L$ near APs, with $L$ a design parameter. Let $d_{m,m'}$ be the 3D distance between the $m$-th and the $m'$-th APs and $O_m \, : \, \{1,\ldots, M \} \rightarrow \{1,\ldots, M \}$ denote the sorting operator for the vector $\left[d_{m,1},\ldots, d_{m,M}\right]^T$, such that $d_{m,O_m(1)} \leq d_{m,O_m(2)} \leq \ldots \leq d_{m,O_m(M)}$. 
Defining the set $$\mathcal{G}_m=\left\lbrace m, O_m(1), \ldots,O_m(L-1)\right\rbrace$$ containing $L$ near APs starting from the $m$-th one, the VCs $\mathcal{M}_1,\ldots, \mathcal{M}_J$ are the \textit{unique} sets in $\left\lbrace \mathcal{G}_m \right \rbrace_{m=1}^M$. Otherwise stated, the VCs $\mathcal{M}_1,\ldots, \mathcal{M}_J$ represent the $J$ combinations of $L$ near APs in the network.

\section{Numerical Results}
The simulation setup for the numerical results is detailed in the following. We consider a square area of 1 km$^2$ wrapped around at the edges to avoid boundary effects. We assume $M=100$ APs with a four-element uniform linear array (ULA) with $\lambda/2$ spacing, i.e., $N_{\rm AP} = 4$. The power spectral density (PSD) of the noise is -174 dBm/Hz and the noise figure at the receiver is 9 dB. The system bandwidth is $W=20$ MHz and the carrier frequency is $f_c=1.9$ GHz. 

With regard to the channels from users to the APs, we consider an urban environment with a high density of buildings and obstacles where all the MSs are in non-line-of-sight (NLOS). The LSF coefficient $\beta_{k,m}$ in dB is modelled as in \cite[Table B.1.2.2.1-1]{3GPP_36814_GUE_model}, i.e.:
\begin{equation}
\beta_{k,m} [\text{dB}]=-36.7\log_{10}(d_{k,m})-22.7-26\log_{10}(f_c)+z_{k,m},
\label{pathloss2_Bjo}
\end{equation}
where $z_{k,m} \sim \mathcal{N}\left( 0, \sigma_{\rm sh}^2\right)$ represents the shadow fading. The shadow fading coefficients from an AP to different users are correlated and follow \cite[Table B.1.2.2.1-4]{3GPP_36814_GUE_model}.
%\begin{equation}
%\mathbb{E}[z_{k,m}z_{j,n}]= \left\lbrace
%\begin{array}{llll}
%&\sigma_{\rm sh}^2 2^{-\frac{\rho_{k,j}}{d_0}} \;, & m=n, \\
%&0 \; ,& m \neq n \, ,
%\end{array} \right.
%\label{shadowing_correlation2_Bjo}
%\end{equation}
%where $\rho_{k,j}$ is the distance between the $k$-th and the $j$-th users, $d_0=9$ m and $\sigma_{\rm sh}^2=4$.

We use fractional power control (FPC) where the transmit power of the $k$-th user can be expressed as 
$
\eta_k^{\rm UL}= \text{min} \left( P_{{\rm max},k}, P_0 \zeta_{k,j}^{-\kappa}\right)\, , 
$
where $P_{{\rm max},k}$ is the maximum $k$-th user transmit power, and $P_0$ is a configurable parameter, $\alpha$ is a path loss compensation factor. Moreover, $\zeta_{k,j}$ captures the LSF that the $k$-th user experiences to the serving APs in the VC $\mathcal{M}_j$, and it is obtained as
$
\zeta_{k,j}= \sqrt{\sum_{m \in \mathcal{M}_j} {\beta_{k,m}}} \, .
$
In the simulations, we use $P_{{\rm max},k}=100$ mW $\forall \; k$, $P_0=-10$ dBmW and $\kappa=0.5$.
Note that for the solution of Problem \eqref{Prob:Matching_problem1} via Algorithm \ref{Hun_method} we need to compute the entries of matrix $\mathbf{F}$ that depends on $\eta_1, \ldots, \eta_K$. For the solution of Problem \eqref{Prob:Matching_problem1}, we assume a uniform power allocation in uplink, i.e., $\eta_k=P_{{\rm max},k}$, $\forall \, k=1,\ldots,K$ and for the evaluation of the performance we use the FPC.
In the following results, we compare the performance of the proposed PBVC with the full-cell-free (FCF) in \cite{ngo2015cell,Ngo_CellFree2017}, in which each user is served by all the APs in the network, with the user-centric (UC) association in \cite{buzzi_CFUC2017,Buzzi_DAndrea_Zappone_TWC2019}, in which each user is served by the $L$ APs that it receives with best average channel conditions, with the large-scale fading decoding (LSFD) in \cite{nayebi2016performance}, where the data from all the APs are weighted at the CPU using only the LSF coefficients, and with the LSFD applied only to the subset of users in the system selected with the proposed PBVC (LSFD+PBVC).
\begin{figure}[!t]
\centering
\includegraphics[scale=0.45]{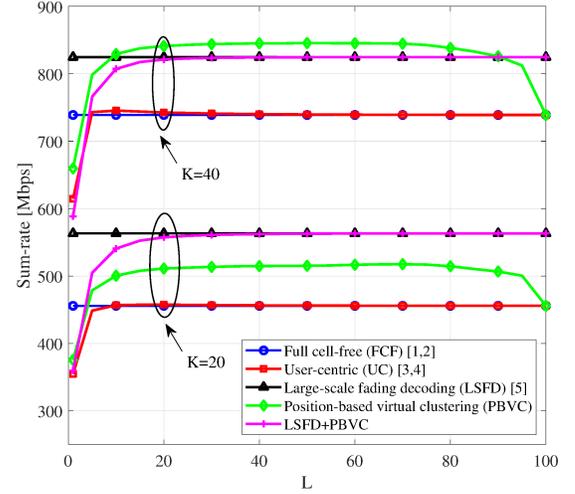}
\caption{Sum-rate versus $L$, i.e., number of APs in each VC, comparison of the proposed PBVC with the FCF, the UC, the LSFD and with the LSFD applied only to the subset of users in the system selected with the PBVC (LSFD+PBVC). Parameters: $M=100$, $N_{\rm AP}=4$ and $\tau_p=16$.}
\label{Fig:SR_vs_L}
\end{figure}

\begin{figure}[!t]
\centering
\includegraphics[scale=0.45]{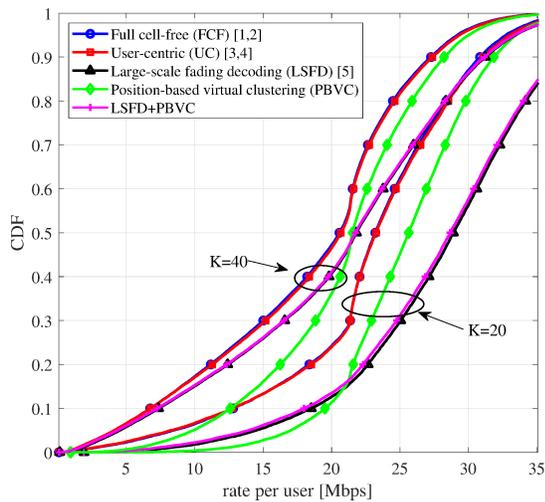}
\caption{CDFs of the rate per user, comparison of the proposed PBVC with the FCF, the UC, the LSFD and with the LSFD applied only to the subset of users in the system selected with the PBVC (LSFD+PBVC). Parameters: $M=100$, $K=30$, $L=20$, $N_{\rm AP}=4$ and  $\tau_p=16$.}
\label{Fig:CDF_Rate}
\end{figure}

Fig. \ref{Fig:SR_vs_L} shows the performance versus $L$, the number of APs in each VC, of the proposed PBVC and of the FCF, UC and LSFD and LSFD+PBVC approaches. The PBVC outperforms both the UC and the FCF confirming that in a cell-free massive MIMO network it is better to decode the users' symbols with the ``right'' subset of APs with respect to decode them by all the APs in the network. Additionally, the proposed approach offers better performance with respect to the LSFD and LSFD+PBVC in terms of sum-rate when the number of users is large, i.e., $K=40$, while the LSFD and LSFD+PBVC outperforms the proposed approach when $K=20$. This can be justified by the fact that the LSFD implements an additional weighting processing at the CPU improving the performance with respect to the maximum-ratio processing but when the number of users is larger, the amount of interference in the system is higher and selecting the right subset of APs in the system reduces the amount of the interference collected at the CPUs. We can see also that the LSFD+PBVC offers the same performance of the LSFD, that uses all the APs in the network, for $L\geq 20$ with a lower load on the cooperation link between the primary CPUs. In Fig.  \ref{Fig:CDF_Rate} we report the cumulative distribution functions (CDFs) of the rate per user in the considered approaches, focusing on $L=20$. We can note that the PBVC offers good performance also in terms of rate per user improving the minimum performance of the users in the system. In particular, the 95\%-likely throughput of the proposed approach when $K=40$ is more than doubled, in fact it increases by about 145\%  with respect to the FCF and UC and by about 120\% with respect to the LSFD. For a lower number of users, i.e., $K=20$, the 95\%-likely throughput of the proposed approach increases by about 112\% with respect to the FCF and UC and by 20\% with respect to the LSFD.

\section{Conclusions}
In this paper, we proposed a user association in a scalable cell-free massive MIMO system where the users are decoded only by a subset of APs in the network. The APs are grouped into cell-centric clusters connected to different cooperative CPUs and each user is associated with a virtual cluster of APs. The proposed user association procedure is aimed at the maximization of the sum-rate of the users in the system. In the numerical simulation, we compare the performance of the proposed technique with other approaches, in particular with the FCF, UC, and LSFD. We can note that the proposed technique does not outperform the alternatives in all the cases, but it offers a considerably lower backhaul load with a negligible performance loss with respect to the FCF and LSFD. Finally, we compare the performance obtained with the combination of the proposed approach with the LSFD, and we note that serving the users with the right set of APs offers the same performance of the LSFD applied to all the APs with a beneficial impact on the total backhaul load of the network.

\section*{Acknowledgment}
The work of C. D'Andrea has been supported by the MIUR Project ``Dipartimenti di Eccellenza 2018-2022'' and by the MIUR PRIN 2017 Project ``LiquidEdge''. The work of E. G. Larsson was partially supported by Swedish research council (VR) and ELLIIT.

\ifCLASSOPTIONcaptionsoff
  \newpage
\fi
\bibliographystyle{IEEEtran}

\bibliography{Cell_free_references}

\end{document}